\documentstyle[prd,aps,graphicx]{revtex}
\begin{document}
\draft
\title{Effective Chiral Theory for Pseudoscalar and Vector Mesons}
\author{ E.Gedalin\thanks{gedal@bgumail.bgu.ac.il},
 A.Moalem\thanks{moalem@bgumail.bgu.ac.il}
 and L.Razdolskaya\thanks{ljuba@bgumail.bgu.ac.il}}
\address{Department of Physics, Ben Gurion University, 84105,Beer Sheva, 
Israel} 
\maketitle
\begin{abstract}We consider the vector meson mixing scheme and mass 
splitting within the framework of an extended $U(3)_L\bigotimes U(3)_R$ 
chiral effective field theory based on the hidden local symmetry 
approach where, the pseudoscalar and vector meson nonets play the 
role of dynamic variables. Unlike other variants of this model, we show 
that the diagonalization of the vector meson mass matrix and the 
assumption that its eigenvalues are identical with the physical meson 
masses, determines the mixing scheme as well as the model free
parameters. We show that VMD can be derived from our lagrangian and that 
for electromagnetic processes at low momenta the VMD model is a good 
first approximation. The model reproduces nicely the radii of the 
charged pions and kaons.
\end{abstract}
\bigskip

\section{Introduction}

In a recent contribution\cite{gedalin01} it was indicated that, for any 
effective field theory (EFT) of colorless meson fields, the mixing 
schemes of particle states and decay constants are determined in 
a unique way by the kinetic and mass lagrangian densities. In a general 
case, these densities are bilinear in terms of the intrinsic fields, 
involving nondiagonal kinetic and mass matrices. However, they can be 
reduced into a standard quadratic form by transforming the intrinsic 
fields into the physical ones in three consecutive steps. These steps 
include : 
(i) the diagonalization of the kinetic matrix,
(ii) rescalling of intrinsic fields to restore the standard normalization
of the kinetic term, and 
(iii) the diagonalization of the resulting mass matrix. 
In such case where the dimensions of the nondiagonal kinetic and
mass submatrices are respectively, $k\times k$ and $n\times n$, this
procedure leads to schemes which involve  $[k(k-1)/2] + [n(n-1)/2]$ mixing
angles and $k$ field rescalling parameters. The commonly used mixing
schemes correspond to specific choice of the kinetic and mass 
matrices. In particular, $\eta-\eta '$ mixing requires one mixing 
angle scheme, if and only if, the kinetic term for the intrinsic fields  
has a quadratic form. Such are the traditional scheme \cite{pdg00}
and the one proposed by Feldmann et al.\cite{feld98,feld98a,feld98b},
the so called quark flavor basis (QFB) scheme.
For bilinear kinetic and mass lagrangian densities (i.e. with nondiagonal
kinetic and mass matrices) mixing schemes involve two mixing angles 
similar to the ones proposed by Escribano and Fr\`{e}re\cite{escrib99}.

There have been several attempts to incorporate within a single effective 
lagrangian a wide range of different electroweak and strong processes
\cite{bando,bramon95,benayoun981,benayoun98}. Particularly
interesting are the Bando, Kugo and Yamawaki (BKY) model\cite{bando},  
based on the  hidden local symmetry (HLS) approach, and variants of this 
model\cite{bramon95,benayoun981}, where the vector mesons 
($\rho , \omega , K^*, \bar{K}^*, \phi $) are identified with the 
dynamical gauge bosons of the HLS  nonlinear chiral lagrangian. 
It was indicated by Benayoun et al.\cite{benayoun981}, that the original
 BKY model neither reproduces the observed vector mass splitting 
nor preserves current conservation for the pseudoscalar octet-singlet 
sector. A variant of this model proposed by Bramon et al.\cite{bramon95} 
also does not predict the masses correctly, although it guarantees 
current conservation, explicitly. Yet another variant of this 
model\cite{benayoun981} includes the $\eta '$ meson also, and to some 
extent cures these deficiencies.  However, once the HLS lagrangian is 
extended to include the $\eta '$ meson, the trace of the pseudoscalar 
(and vector) nonet field matrix no longer vanishes, and therefore, a 
general expression for the lagrangian must include additional terms 
which are quadratic in the traces of the nonet field matrices. More 
recently, a most general lagrangian based on local 
$U(3)_L\bigotimes U(3)_R$ 
symmetry was constructed \cite{gedalin00,gedalin011},
by combining the HLS approach with a general procedure of  including the
$\eta'$ meson into a chiral theory\cite{gasser85,leut96,leut97,herera97}.
Both the symmetric parts of the lagrangian, as well as the symmetry 
breaking companions, include terms quadratic in the traces of the
field matrices. Likewise, the lagrangian includes an explicit 
$U(3)_L\bigotimes U(3)_R$ symmetry violating mass term for the pseudoscalar 
mesons\cite{leut96,herera97}. These additional terms influence the 
meson mass spectra and more importantly, by means of the 
diagonalization procedure of Ref.\cite{gedalin01}, not only the mixing 
scheme is defined uniquely but also, taking the eigenvalues of the 
resulting mass matrix to be equal to the physical masses, it was possible 
to fix a number of the  model free parameters. All observables are 
calculated from the lagrangian straightforwardly without additional 
assumptions. Albeit, the lagrangian constructed reflects the fundamental
symmetries of the QCD lagrangian and provides a coherent theoretical
framework where the  pseudoscalar and vector meson nonets
are treated on equal footing. 

Our main interest in the present work is to further develop and explore
the $U(3)_L\bigotimes U(3)_R$ chiral theory of Ref.\cite{gedalin011} 
by considering the vector meson sector. It is demonstrated that in this 
case the relevant model parameters can all be fixed through the
diagonalization procedure of Ref.\cite{gedalin01}. It has been illustrated 
already\cite{bando} that the vector meson dominance (VMD) model emerges 
as a natural consequence of the basic assumption of the HLS approach where 
vector mesons play the role of gauge bosons. Particularly, the VMD model 
becomes a dynamical consequence for a specific choice of the HLS model 
parameter $a = 2$. Detailed analyses of data argue for departures from 
this value\cite{benayoun981,gao00}, and that in addition to the vector 
meson poles in the charged meson form factors, a direct photon-
pseudoscalar meson coupling may also contribute significantly. 
More recently, Harada and Yamawaki\cite{harada01} have shown that in 
an effective field theory based on HLS, VMD may be badly violated. 
We demonstrate in this work that VMD derived from the lagrangian 
constructed in Ref.\cite{gedalin011} is a good first approximation 
for electromagnetic processes at low momenta. Our paper is organized 
as follows. In Section II we present briefly the lagrangian. In 
section III we discuss vector mixing scheme and mass splitting 
corresponding to the lagrangian presumed in the simple case of exact 
isospin symmetry and in the more evolved case of isospin symmetry 
breaking. In Section IV we consider the photon interaction with hadronic 
matter and discuss the relation between the lagrangian constructed 
and the traditional VMD model. We summarize and conclude in Section V.
  
\bigskip

\section{The effective lagrangian}

As in Ref.\cite{gedalin011}, the lagrangian is written in the form,
\begin{equation}
        L = L_A + \bar{L}_A + L_m + a(L_V + \bar{L}_V) -
         \frac{1}{4}Tr(V_{\mu\nu}V^{\mu\nu}) + L_{WZW} + \ldots ~,
        \label{elag}
\end{equation}
where $L_A (\bar{L}_A)$ and $ L_V (\bar{L}_V)$ are the symmetric parts
(symmetry breaking companions) for the pseudoscalar  and vector 
mesons,  $L_m$ is a most general $U(3)_L\bigotimes U(3)_R$ symmetry
violating mass term\cite{leut96,herera97}, $L_{WZW}$ stands for the well 
known Wess-Zumino-Witten term \cite{witten83,callan84}, and the ellipsis 
"$\ldots$"  represents terms accounting for the regularization of 
one loop contributions \cite{bijnens90,bijnens901,gasser85}. All terms 
in the expression above are constructed from boson gauge fields $V_\mu$, 
and vector and axial-vector covariants defined 
as\cite{gasser88,krause90,bernard95},
\begin{eqnarray}
        && \Gamma_\mu =
   \frac{i}{2}\left[\xi^\dagger ,\partial_\mu\xi\right]
          +\frac{1}{2}
\left(\xi^\dagger r_\mu \xi + \xi l_\mu\xi^\dagger\right) ~,
        \label{gammac} \\
        &&\Delta_{\mu} =
     \frac{i}{2}\left\{\xi^\dagger ,\partial_{\mu}\xi\right\}
  +\frac{1}{2}\left(\xi^\dagger r_\mu\xi - \xi l_\mu\xi^\dagger\right) ~.
        \label{deltaf}
    \end{eqnarray}
Here $r_\mu$ and $l_\mu$ represent the standard model external gauge
fields; $r_\mu= v_\mu + a_\mu$ and $ \quad l_\mu = v_\mu -
a_\mu$, with $v_\mu$ and  $a_\mu$ being the vector and axial
vector external electroweak fields, respectively. The non-linear
representation of the pseudoscalar nonet fields is taken as
\cite{gasser85,leut96,herera97},   
\begin{equation}
        U(P,\eta_0+F_0\vartheta) \equiv \xi^2(P,\eta_0+F_0\vartheta)
        \equiv \exp 
\left\{i\frac{\sqrt{2}}{F_8}P + i\sqrt{\frac{2}{3}}\frac{1}{F_0}\eta_0 
{\bf 1}\right\}~,
        \label{pmfield}
\end{equation}
with $\eta_0(x)$ being the pseudoscalar singlet and $P(x)$ the
pseudoscalar Goldstone octet matrix,
 \begin{equation}
   P = \left(
\begin{array}{ccc}
\frac{1}{\sqrt{2}}\pi^0+\frac{1}{\sqrt{6}}\eta_8 & \pi^+ &K^+ \\
        \pi^- & - \frac{1}{\sqrt{2}}\pi^0+\frac{1}{\sqrt{6}}\eta_8
        &K^0  \\
        K^- & \bar{K}^0 &-\frac{2}{\sqrt{6}}\eta_8
        \end{array}
      \right)~,
        \label{poctet}
        \end{equation}
with obvious notations. The vacuum angle $\vartheta (x)$ is an auxiliary
field which renders the variable $X (x)\equiv \sqrt{6}\eta_0(x)/F_0 + 
\vartheta (x)$ to be invariant under $U(3)_L\bigotimes U(3)_R$ 
transformations\cite{gasser88}. In this representation, the 
unimodular part of $U$ contains the octet degrees of freedom while 
the phase  $detU = \exp (i\sqrt{6}\eta_0/F_0)$ involves the 
singlet only. To lowest order (i.e., smallest number of derivatives) the 
lagrangian is built from traces of the covariants $\Delta_{\mu}$, 
$\Gamma_{\mu} - gV_{\mu}$, $\Delta_{\mu}\Delta^{\mu}$, 
$(\Gamma_{\mu} - gV_{\mu})(\Gamma^{\mu} - g V^{\mu})$, 
$D^\mu \vartheta$, and arbitrary functions of the variable $X(x)$, 
all being invariant under $U(3)_L\bigotimes U(3)_R$ transformations. 
Explicit expressions of various lagrangian parts are listed for 
convenience in the Appendix. Their derivation can be found in Ref.
\cite{gedalin011}. Although the lagrangian, Eqn.\ref{elag} appears 
similar to that of Bando et al. \cite{bando} and other variants of this 
model\cite{bramon95,benayoun981}, the terms $L_i$ and their symmetry
 breaking companions $\bar{L}_i (i = A,V)$ are different, including 
additional terms proportional to $Tr (\Delta_{\mu})Tr (\Delta^{\mu})$, 
$Tr (\Delta_\mu) D^{\mu} \vartheta$ and $Tr(\Gamma^{\mu} - g V^{\mu}) 
Tr(\Gamma_{\mu} - g V_{\mu})$. Clearly, such terms must be included in any 
$U(3)_L\bigotimes U(3)_R$ symmetry based HLS theory, where the traces of
$\Delta_\mu$ and $\Gamma_\mu - g V_\mu $ do not vanish. It is demonstrated 
below that these terms are needed in order to reproduce the meson masses 
from the lagrangian without imposing additional constraints. We add also 
that the matrix used to generate the symmetry breaking companions is a 
universal Hermitian matrix defined as,
\begin{equation}
B = \xi b \xi + \xi^\dagger b \xi^\dagger~;~~~~ b= diag(b_u, b_d, b_s)~,
\end{equation}
where $b_i = m_i/(m_u + m_d + m_s)$. This form guarantees that the ratio of
$SU(3)_F$ symmetry breaking scales to isospin symmetry breaking scales be 
identical to the once of the underlying QCD.  Note that, the 
lagrangian\ref{elag} involves coefficient functions $W_i(X)$ which 
are absent in the $SU(3)$ symmetry limit. Again such functions must be 
included in the process of extending the $SU(3)_L\bigotimes SU(3)_R$ 
based model into a general $U(3)_L\bigotimes U(3)_R$ theory. As a 
final remark we note that the symmetry violating mass term in  Eqn. 
\ref{elag} has the exact form derived by Herera-Sikl\'{o}dy  et 
al.\cite{herera97}. As in Refs.\cite{bando,bramon95,benayoun981}, the sum 
$a(L_V + \bar{L}_V)$ contains, amongst other contributions, a vector meson 
mass term $\sim V_\mu V^\mu$, a vector-photon conversion factor 
$\sim V_\mu A^\mu$ and the coupling of pseudoscalar pairs to both 
vectors and photons. The latter coupling can be eliminated by choosing 
$a=2$ allowing to incorporate the $strict$ VMD lagrangian
\cite{bando,oconnell97}. We shall attempt to determine the value of this 
parameter from data in order to assess the validity of the traditional VMD
model. Altogether, the lagrangian Eqn. \ref{elag} (see Appendix), involves 
symmetry breaking scales $c_i, d_i$ (i=A,V), a general coupling constant 
$g$ and a free parameter $a$. The symmetry breaking scales 
$c_A, d_A ~\mbox{and}~ r$ are determined from global fit analysis of 
radiative decay widths of pseudoscalar and vector mesons\cite{gedalin011}. 
The other symmetry breaking scales $c_V~, d_V ~\mbox{and}~ ag^2$ are 
determined through the diagonalization of the vector mass matrix.

\section{Vector meson masses and mixing} 

We now turn to consider mixing and masses of vector mesons. The vector 
part $a(L_V +\bar{L}_V)$ of the lagrangian \ref{lvbreak} 
(See Appendix) involves a vector meson mass term $\sim V_\mu V^\mu$ 
which determines as already indicated the vector mixing scheme, uniquely. 
Generally speaking, the vector nonet  matrix $V$ is the sum of an 
octet matrix $V_8$, involving the 
$\rho^0, \rho^{\pm}, K^{*0}, \bar{K}^{*0}, K^{*\pm} ~\mbox{and}~\omega_8$ 
meson fields, and an $SU(3)_F$ singlet $\omega_0$ matrix, e.g.,
\begin{equation}
	V = V_8 + r_V \frac{1}{\sqrt{3}} \omega_0 \times 1~.
	\label{vnonet}
\end{equation}  
Here a departure of the parameter $r_V$ from one accounts for possible 
nonet symmetry breaking. However, unlike the case of pseudoscalar mesons, 
the light vector meson spectrum shows no clear evidence for such a 
departure. In the discussion to follow we take $r_V=1$, assuming exact 
nonet symmetry. Then, the vector meson nonet matrix has the form 
(for brevity the Lorentz indices of the vector fields are omitted),
\begin{equation}
     V = \left(\begin{array}{ccc}
    \frac{\rho^0}{\sqrt{2}}+\frac{\omega_8}{\sqrt{6}}+
    \frac{\omega_0}{\sqrt{3}} & \rho^+ & K^{*+}  \\
    \rho^- & - \frac{\rho^0}{\sqrt{2}}+\frac{\omega_8}{\sqrt{6}}+
    \frac{\omega_0}{\sqrt{3}} & K^{*0}  \\
    K^{*-} & \bar{K}^{*0} &-\frac{2\omega_8}{\sqrt{6}}+
    \frac{\omega_0}{\sqrt{3}}
    \end{array}\right)~.
    \label{vnonetm}
    \end{equation}   
As was already indicated in the introduction, the kinetic and mass terms
of the lagrangian can be reduced to a standard quadratic form by first
diagonalizing the kinetic matrix via a unitary transformation $\Upsilon$, 
rescale the fields through a transformation $R$ to restore the standard 
normalization of the kinetic term, and finally diagonalizing the resulting 
mass matrix through another unitary transformation $\Omega$. These three 
steps are sufficient to define the state mixing 
scheme\cite{gedalin01}. 

From Eqns. \ref{vnonetm}, \ref{laginv} and \ref{lvbreak} the vector meson 
kinetic and mass terms amounts to,
\begin{equation}
  L_{km} =  -\frac{1}{4}
(\partial_\mu{V_{\nu}}-\partial_{\nu}{V_{\mu}})
(\partial^\mu{V^{\nu}}-\partial^{\nu}{V^{\mu}})   
           + \frac{1}{2}(V_\mu {\cal M}^2_V V^{\mu}) ~,
\label{kmlag}   
\end{equation}
with,
\begin{eqnarray}
   & & \frac{1}{2}V_\mu{\cal M}^2_V V^{\mu} =
        aF^2_8g^2\left[Tr(V_{\mu} V^{\mu}) 
	+\tilde{w}_4 Tr(V_\mu)Tr(V^\mu)\right. +
       \nonumber \\
   & & \left.c_V\left(Tr(BV_\mu V^\mu)  +
       \tilde{w}_{4} Tr(B V_{\mu}) Tr(V^{\mu})\right) +
       d_V \left( Tr(BV_{\mu} BV^{\mu} )    + 
       \tilde{w}_{4} Tr(BV_{\mu} )Tr(BV^{\mu} ) \right)\right]~.
       \label{mlagv}
\end{eqnarray}
Thus the kinetic term has the standard normalized quadratic form, and  
following \cite{gedalin01} the mixing scheme is determined through the 
diagonalization of the mass matrix only and involves one mixing angle.  
The nonvanishing elements of the vector meson mass matrix read, 
\begin{mathletters}
\begin{eqnarray}
	 &  & \mu^2(\rho^0\rho^0) = \mu^2_V\left(1 + \frac{c_V}{2}(b_u + b_d) + 
	 \frac{d_V}{2}[b^2_u + b^2_d + \tilde{w}_4(b_u-b_d)^2] \right)~, 
    \label{} \\
	 &  & \mu^2(\rho^+\rho^+) = \mu^2_V\left(1 + \frac{c_V}{2}(b_u +b_d) + 
	 d_Vb_u b_d \right)~, 
	\label{} \\
	 &  & \mu^2(\omega_8\omega_8) =
	  \mu^2_V\left(1 + \frac{c_V}{6}(b_u + b_d + 4b_s) + 
	 \frac{d_V}{6}[(b^2_u + b^2_d + 4b^2_s) + 
	 \tilde{w}_4(b_u + b_d -2b_s)^2] \right)~,
	\label{} \\
	 &  & \mu^2(\omega_0\omega_0) =  \mu^2_V\left(1 + 3\tilde{w}_4+
	\frac{c_V}{3}(b_u + b_d + b_s)(1 + 3\tilde{w}_4) + 
	 \frac{d_V}{3}[b^2_u + b^2_d + b^2_s + 
	 \tilde{w}_4(b_u + b_d + b_s)^2] \right)~,
	\label{} \\
	 &  & \mu^2(\rho^0\omega_8) = 
	   \mu^2_V 
	 \frac{1}{2\sqrt{3}}(b_u - b_d)\left(c_V +
	 d_V[b_u + b_d  + \tilde{w}_4(b_u + b_d - 2b_s)]\right)~,
	\label{} \\
	 &  & \mu^2(\rho^0\omega_0) =
	   \mu^2_V \sqrt{\frac{1}{6}}(b_u - b_d)
	\left(c_V(1 + \frac{3}{2}\tilde{w}_4) + d_V[b_u + b_d  + 
	 \tilde{w}_4(b_u + b_d + b_s)]\right)~,
	\label{} \\
	 &  & \mu^2(\omega_8\omega_0) = 
	   \mu^2_V \frac{1}{3\sqrt{2}}
	\left(c_V(b_u + b_d -2b_s)(1 +\frac{3}{2} \tilde{w}_4) + 
	 d_V[b^2_u + b^2_d - 2b^2_s + 
	 \tilde{w}_4(b_u + b_d -2b_s)(b_u + b_d + b_s)] \right)~,
	\label{} \\
	 &  & \mu^2(K^{*\pm}K^{\pm}) =  \mu^2_V\left(1 + \frac{c_V}{2}(b_u +b_s) 
    + \frac{d_V}{2}b_u b_s \right)~,
	\label{} \\
	 &  & \mu^2(K^{*0}K^{*0}) = \mu^2(\bar{K}^0\bar{K}^0)  = 
         \mu^2_V\left(1+ \frac{c_V}{2}(b_d + b_s) + \frac{d_V}{2}b_d b_s 
     \right)~,
	\label{massvec}
\end{eqnarray}
\end{mathletters}
with,
\begin{equation}
	  \mu^2_V = 2ag^2 F^2_8~, 
	\label{} 
\end{equation}
We stress here that even in the limits of exact chiral symmetry 
and nonet symmetry, i.e. $c_V = d_V = 0, ~r_V =1$, the singlet vector
meson mass,  $m _{\omega0} = \mu_V\sqrt{1+3\tilde{w}_4}$, differs from
that of the octet $m_{\omega8} = \mu_V$. This distinct property occurs due
to the term $\tilde{W}_{4}(X) Tr(\Gamma_\mu - gV_\mu)Tr(\Gamma^\mu -
gV^\mu)$ in the lagrangian Eqn. \ref{elag}. Only for $\tilde{w}_4 = 0$ 
(corresponding to lagrangians previously used) all nonet meson 
masses are equal. \footnote{In case of {\it broken\/} nonet
symmetry it is still possible to maintain $m_{\omega 0} \neq \mu_V$  
but with the nonet symmetry breaking parameter fixed
arbitrary to be $r^2_V = 1/(1+3\tilde{w}_4)$}. 
Clearly, the mass matrix is nondiagonal due to the $\omega_8-\omega_0$, 
$\rho^0-\omega_8$ and $\rho^0-\omega_0$ 
mixing.                                  
Notice though that the  $\rho^0 -\omega_8$ and 
$\rho^0-\omega_0$ admixtures are extremely small and 
and vanish in the limit of exact isospin symmetry ($b_u = b_d$). 

First we diagonalize the mass matrix in the limit of exact
isospin symmetry. In this limit, taking $b_s = 1$ and neglecting 
terms of the order of $c_V b_i, d_V b_i (i = u,d)$, the matrix mass 
assumes a particularly simple form with elements,
\begin{mathletters}
\begin{eqnarray}
         &  & \mu^2(\rho\rho) = \mu^2_V~,
        \label{rhom} \\
         &  & \mu^2(\omega_8\omega_8) =  \mu^2_V\left(1 + \frac{2c_V}{3} +
         \frac{2d_V}{3}(1 + \tilde{w}_4) \right)~,
        \label{omega8m} \\\
         &  & \mu^2(\omega_0\omega_0) =
         \mu^2_V\left((1 +\frac{c_V}{3})(1 + 3\tilde{w}_4) +
         \frac{d_V}{3}(1 + \tilde{w}_4) \right)~,
        \label{omega0m} \\
         &  & \mu^2(\omega_8\omega_0) =  - \mu^2_V\frac{\sqrt{2}}{3}
        \left(c_V(1 +\frac{3}{2} \tilde{w}_4) 
         + d_V(1 + \tilde{w}_4) \right)~,
        \label{omega80m} \\
         &  & \mu^2(K^{*}) =  \mu^2_V\left(1 + \frac{c_V}{2} \right)~, 
        \label{kaonm} 
\end{eqnarray}
\end{mathletters}
which we  diagonalize using the unitary transformation,
\begin{equation}   
        \left(
        \begin{array}{c}
                \omega_8  \\
                \omega_0
        \end{array}
        \right) = \left(
        \begin{array}{cc}
                \cos\theta_V & \sin\theta_V \\
                -\sin\theta_V & \cos\theta_V
        \end{array}
        \right)   \left(
        \begin{array}{c}
                \omega  \\
                \phi
        \end{array}
        \right)~.
        \label{omega}
\end{equation}
It is straightforward to show that the resulting  physical meson masses 
and mixing angle are,
\begin{eqnarray}
   &  & m^2_{\omega} = \frac{1}{2}\left(\mu^2(\omega_8\omega_8)
   +\mu^2(\omega_0\omega_0) -
   \sqrt{(\mu^2(\omega_0\omega_0)- \mu^2(\omega_8\omega_8))^2 +
   4(\mu^2(\omega_0\omega_8))^2}\right) ~,
   \label{omegam} \\
   &  &   m^2_{\phi} =  \frac{1}{2}\left(\mu^2(\omega_8\omega_8)
   +\mu^2(\omega_0\omega_0) +
   \sqrt{(\mu^2(\omega_0\omega_0)- \mu^2(\omega_8\omega_8))^2 +
   4(\mu^2(\omega_0\omega_8))^2}\right) ~,
   \label{phim} \\
   &  &\tan 2\theta_V
   = - \frac{ 2\mu^2(\omega_0\omega_8)}
   {(\mu^2(\omega_0 \omega_0) - \mu^2(\omega_8 \omega_8)}
   =-\sqrt{ \frac{(m^2_\phi-m^2_\omega)^2}
   {(m^2_\phi+m^2_\omega - 2 \mu^2(\omega_8 \omega_8)^2} - 1}~.
   \label{mixinga}
\end{eqnarray}
From Eqns. \ref{rhom}-\ref{kaonm}, with $\tilde{w}_4 = 0$ the mixing angle 
reduces to the ideal mixing angle value $\theta_{Videal}=-54.7^o$, and 
more importantly does not depend on the symmetry breaking scales 
$c_V~\mbox{and}~d_V$. 
Correspondingly, the $\omega,~K$ and $\phi$ meson masses become,
\begin{eqnarray}
        m^2_{\omega} = m^2_\rho ~, & ~~ m^2_K = m^2_\rho (1 +
\frac{1}{2}c_V)~,
 & ~~ m^2_{\phi} = m^2_\rho(1+c_V+d_V)~.
        \label{mesmass}
\end{eqnarray}
Clearly, assuming now $d_V= c^2_V/4$ leads to the BKY\cite{bando} mass 
relations,  
\begin{equation}
        \frac{m^2_K}{m^2_\rho} = \frac{m^2_\phi}{m^2_K} = 1 +
\frac{1}{2}c_V~.
        \label{bkyrel} 
\end{equation} 
Otherwise, taking $d_V=0$ leads to the mass relations of Bramon et al.
\cite{bramon95}
\begin{equation}
\quad \frac{m^2_K}{m^2_\rho} = 1 + \frac{1}{2}c_V~,
        \quad \frac{m^2_{\phi}}{m^2_\rho} = 1+c_V~.
        \label{brrel}
\end{equation}
It should be stress though that in all of these private cases, the
$\rho$ and $\omega$ masses are equal and $\theta_V = \theta_{Videal}$. To 
describe the actual vector meson masses we should keep $d_V$ as a free 
parameter and  $\tilde{w}_4 \neq 0$. Based on the proximity of the 
$\rho$ and $\omega$ masses, the parameter $\tilde{w}_4$ should be very 
small, and to first order the $\omega~\mbox{and}~\phi$ meson masses are,
\begin{equation}
        m^2_\omega =m^2_\rho (1+2\tilde{w}_4)~,~~~
        m^2_\phi= m^2_\rho(1+c_V+d_V)(1+\tilde{w}_4)~.
        \label{gmrrel}
\end{equation}
Thus, the parameter  $\tilde{w}_4$ determines the $\rho -\omega$ mass 
splitting
while $(c_V + d_V) = 2(m^2_\phi - m^2_\rho)/(m^2_\rho + m^2_\omega)$.
At this order, using the expressions above and Eq.\ref{mesmass} for 
$m^2_K$, along with the experimental meson mass values\cite{pdg00} 
one obtains, $c_V = 0.69\pm 0.01,~ d_V \approx 0.03$ and 
$\tilde{w}_4 \approx 0.01$. Indeed, $\tilde{w}_4 $ is very small
 but it plays an important role in reproducing the $\rho-\omega$ mass 
 splitting. Note also that the actual value of the parameter $d_V$ is 
 far smaller than $c^2_V/4 = 0.12$ favoring the Bramon et al.\cite{bramon95}
  relations Eqn.\ref{brrel}. More accurate 
values of these parameters can be deduced from the exact expressions for 
the vector meson masses. These are, $\tilde{w}_4 = 0.009\pm 0.0001 
~~\mbox{and}~~ d_V = 0.025\pm 0.01$. Correspondingly, the mixing angle 
for the vector meson states is $\theta_V=-(52\pm 1)^o$, rather close to
ideal mixing. In turn, this value constrains the field admixtures due to 
nonideal mixing to be very small. With nonideal mixing taken into account, 
the matrix of the physical vector fields reads,
\begin{equation}
     V = \left(\begin{array}{ccc}
     \frac{\rho^0+\omega+\epsilon\phi}{\sqrt{2}}
     & \rho^+ & K^{*+}  \\
     \rho^- & \frac{-\rho^0+\omega+\epsilon\phi}{\sqrt{2}} 
     & K^{*0}  \\
     K^{*-} & \bar{K}^{*0} &\phi -\epsilon \omega
     \end{array}\right)~.
     \label{vnonet}
\end{equation} 
where $\epsilon = 0.047\pm 0.001$, being a measure of a non-strange 
(strange) admixture in the physical $\phi ~\mbox{and} ~\omega)$ fields. 
We stress again that this parameter like the symmetry breaking scales 
$c_V, d_V$ and ${\tilde \omega}_4$ are all determined through the 
diagonalization of the mass matrix and the requirement that the 
eigenvalues of this matrix should reproduce the physical masses.

We now consider the effects of isospin symmetry breaking on the mass 
splitting and mixing angle. The nondiagonal part of the mass matrix
 now reads
\begin{equation}
	{\cal M}^2(\rho^0,\omega_8,\omega_0) = \left(
	\begin{array}{ccc}
		\mu^2(\rho^0\rho^0) & \mu^2(\rho^0\omega_8) & \mu^2(\rho^0\omega_0) \\
		\mu^2(\rho^0\omega_8) & \mu^2(\omega_8\omega_8) 
		& \mu^2(\omega_8\omega_0)  \\
		\mu^2(\rho^0\omega_0) & \mu^2(\omega_8\omega_0) 
		& \mu^2(\omega_0\omega_0)
	\end{array}
	\right)
	\label{}
\end{equation}
It is convenient to write the transformation $\Omega$ which diagonalize 
this submatrix in the form
\begin{equation}
	\Omega = \Omega(\omega_8,\omega_0)\Omega(\rho^0, \omega, \phi)
	\label{€}
\end{equation}
where the transformation $\Omega(\omega_8,\omega_0)$ diagonalize the 
$\omega_8, \omega_0$ submatrix 
\begin{equation}
	\Omega(\omega_8,\omega_0) =\left(
	\begin{array}{ccc}
		1 & 0 & 0  \\
		0 & \cos \theta_V & \sin\theta_V  \\
		0 & -\sin\theta_V & \cos\theta_V
	\end{array}
	\right)
	\label{}
\end{equation}
and the transformation $\Omega(\rho^0, \omega, \phi)$ diagonalize the 
reduced mass matrix
\begin{equation}
     \Omega^{-1}(\omega_8,\omega_0){\cal 
M}^2(\rho^0,\omega_8,\omega_0)                     
     \Omega(\omega_8,\omega_0) =\left(\begin{array}{ccc}
	m^2_\rho & m^2_{\rho\omega} & m^2_{\rho\phi}  \\
	m^2_{\rho\omega} & m^2_\omega & 0  \\
	m^2_{\rho\phi} & 0 & m^2_\phi
	\end{array}\right)~,
	\label{redmmat}
\end{equation}  
where,
\begin{equation}
        m^2_{\rho\omega}=\frac {1}{2}m^2_\rho (b_u-b_d)c_V~,~~~~
        m^2_{\rho\phi}  =\frac{1}{2\sqrt{2}}m^2_\rho 
(b_u-b_d)c_V\tilde{w}_4~. 
        \label{}
\end{equation}
From the Eq.\ref{redmmat}it is clear that effects of isospin symmetry 
breaking 
on the $\rho ~\mbox{and}~\omega$ masses are negligibly small $\approx 
0.1\%$
Taking $\tilde{w}_4 = 0.009$ as above, it is rather easy to show that 
the matrix element $m^2_{\rho\phi}$ is negligibly small and consequently
the $\rho^0-\phi$ mixing angle is practically zero. Therefore we may 
write the $\Omega(\rho^0, \omega, \phi)$  transformation in the simple form
\begin{equation}
	\Omega(\rho^0, \omega, \phi) = \left(
	\begin{array}{ccc}
		\cos \theta_{\rho\omega} & \sin \theta_{\rho\omega} & 0  \\
		-\sin \theta_{\rho\omega} & \cos \theta_{\rho\omega} & 0  \\
		0 & 0 & 1
	\end{array}\right)
	\label{}
\end{equation}
where the $\rho-\omega$ mixing angle is given by the expression,
\begin{equation}
	\tan 2\theta_{\rho\omega} 
=-2\frac{m^2_{\rho\omega}}{m^2_\rho-m^2_\omega}~.
	\label{}
\end{equation}
With this in mind, we may now write the vector meson matrix in the form,
\begin{equation}
     V = \left(\begin{array}{ccc}
		\frac{X_1\rho^0+X_2\omega+\epsilon\phi}{\sqrt{2}}
		& \rho^+ & K^{*+}  \\
		\rho^- & \frac{-X_2\rho^0+X_1\omega+\epsilon\phi}{\sqrt{2}} 
		  & K^{*0}  \\
		K^{*-} & \bar{K}^{*0} &\phi +\epsilon(\rho^0\sin\theta_{\rho\omega}
		- \omega\cos\theta_{\rho\omega})
	\end{array}\right)~,
	\label{vnonet}
	\end{equation} 
with,
	\begin{eqnarray}
	 &  & X_1 =\cos \theta_{\rho\omega} - \sin \theta_{\rho\omega}~,
	\label{} \\
	 &  & X_2 =\cos \theta_{\rho\omega} + \sin \theta_{\rho\omega}.
	\label{}
	\end{eqnarray} 

To evaluate the $\rho^0-\omega$ mixing angle, we need an estimate 
of the quark masses ratios $b_i, (i= u,d,s)$. We recall that,
\begin{eqnarray}
         &  & m^2_{K^+} = \frac{1}{2(b_u + b_d)}m^2_\pi(b_u +
b_s)(1+\frac{1}{2}c_A)~,
        \label{} \\
         &  &  m^2_{K^0} = \frac{1}{2(b_u + b_d)}m^2_\pi(b_d +
b_s)(1+\frac{1}{2}c_A)~.
        \label{}
\end{eqnarray}
Then, we may estimate $(b_u - b_d)$ from the kaon masses\cite{pdg00} 
according to,
\begin{equation}
	\frac{b_u - b_d}{1-(b_u +b_d)/2} = 2\frac{ m^2_{K^+} -  m^2_{K^0}}
	{ m^2_{K^+} +  m^2_{K^0}}~,
	\label{}
\end{equation}
which yields,
\begin{equation}
	b_u - b_d = -0.0160 \pm 0.0005~.
	\label{}
\end{equation}
With this value of $b_u - b_d$ the calculated $\rho^0-\omega$ mixing 
angle  reads,
\begin{equation}
	\theta_{\rho\omega} = (-8.9 \pm 0.3)^o~.
	\label{}
\end{equation}
For the parameter $b_s$ we have\cite{gedalin011},
\begin{equation}
 \frac{m_s}{m} = \frac{2m_s}{m_u +m_d}= \frac{2b_s}{1-b_s} = 
 2\left(1+\frac{1}{2}c_A\right)\left(\frac{m_K}{m_\pi}\right)^2 -1~,
 \label{}
\end{equation}
which yields for $c_A=0.64\pm 0.06$ ({\it Alternative 1\/} of 
Ref.\cite{gedalin011}) 
\begin{equation}
 b_s = 0.940 \pm 0.01~,
 \label{}
\end{equation}
and consequently $b_u = 0.022\pm 0.005,~~b_d = 0.038\pm0.005$. For the 
{\it Alternative II\/} ($c_A = 0.2\pm 0.05$) an QFB mixing scheme ($c_A = 
1.4\pm 0.1$) we obtain $b_s=0.928,~ b_u=0.028,~ b_d=0.044$ and 
$b_s=0.953,~b_u = 0.0155,~b_d = 0.0315$ respectively.
This complete defining the mixing scheme in the case of isospin symmetry 
breaking. Here as in the analysis of the pseudoscalar meson sector
\cite{gedalin011}, the exact form of the matrix B used to generate 
flavor symmetry breaking (see Appendix), guarantees that the ratios 
of $SU(3)$ symmetry breaking scales to isospin symmetry breaking scales 
are identical to the ones of the underlying QCD. 

\section{Vector meson dominance model and pseudoscalar meson form factors}

A particularly interesting aspect of the theory presented above is the 
interaction between the photon and hadronic matter. It was already 
illustrated by Bando et al.\cite{bando} and O'Connell et al.
\cite{oconnell97} that the HLS lagrangian is closely related to the well 
known VMD type 2 lagrangian and that by suitable transformation it can 
be reduced into the VMD type 1 lagrangian. (See also Ref.\cite{benayoun981} 
on this matter). The lagrangian, Eqn.\ref{elag}, contains terms 
proportional to $\sim V_\mu V^\mu ~~\mbox{and}~~A_\mu A^\mu$ and 
$\sim V_\mu A^\mu$, corresponding respectively to a vector meson and 
photon mass terms and photon-vector meson interactions. likewise, it 
contains the coupling of both vector mesons and photons to pairs of 
pseudoscalar mesons. By following arguments similar to those of 
Ref.\cite{oconnell97} we derive in this section explicit expressions 
for the photon, vector, and pseudoscalar parts of the lagrangian,
 Eqn.\ref{elag}. We recall first that the covariants 
$\Gamma_\mu$ and $\Delta_\mu$, Eqns.\ref{gammac},\ref{deltaf} can be 
expanded as,
\begin{eqnarray}
    & & \Gamma_\mu = -eA_\mu Q + \frac{i}{4F^2_\pi}\left[{\cal P},
    \partial_\mu{\cal P}\right] -\frac{ie}{2F^2_\pi}A_\mu
    \left(\{{\cal P}^2, Q\} - {\cal P}Q{\cal P}\right) + \ldots ~, 
        \label{gammac1}\\
    & & \Delta_\mu = -\frac{1}{F_\pi \sqrt{2}}\partial_\mu{\cal P} +
     i\frac{e}{F_\pi \sqrt{2}}A_\mu [Q,{\cal P}] + \ldots~,
        \label{deltaf1}
\end{eqnarray}
where $A_\mu$ stands for the photon field, $ Q = diag (2/3, -1/3, -1/3)$
the quark charge operator, and $ \cal P$ is the pseudoscalar nonet matrix
expressed in terms of the physical fields (see Ref. \cite{gedalin011}). 
The ellipsis "$\ldots$" denotes terms involving higher powers of the 
fields. The covariant $\Gamma_\mu -gV_\mu$ contains then terms
proportional to ($eA_\mu Q - gV_\mu$). 
\footnote{Strictly speaking, $\Gamma_\mu -gV_\mu$ involves the terms 
$-e(A_\mu -\tan \theta_W Z^0_\mu)Q 
-\frac{e}{2\sin\vartheta_W\cos\vartheta_W}T_z  
Z^0_\mu - \frac{e}{2\sqrt{2}\sin\vartheta_W}W_\mu -gV_\mu$ which give rise 
to direct $V-A,~V-Z^0~\mbox{and}~V-W$ couplings.} 
By substituting the expression, Eqn.\ref{gammac1} into the lagrangian part 
$L_V + \bar{L}_V$ we obtain the following photon and vector meson terms : 
\\
(i) a direct photon-vector meson coupling,
\begin{eqnarray}
 & & aL(V,\gamma) = aF^2_8geA^\mu\left[Tr\left(\{V_\mu,Q\}(1+c_VB)\right) +
 \right.
 \nonumber \\ 
 & &\left.c_V Tr(V_\mu)Tr(BQ) +d_V Tr \left(\{BV_\mu, BQ\}\right) + 
          2d_VTr(BQ)Tr(bV_\mu)\right]~,
 \label{lagvph}
 \end{eqnarray}
(ii) a  photon mass term,
\begin{equation}
	aL_{m\gamma} = a\frac{F^2_8e^2}{3}A^2_\mu \left(2 +\frac{1}{3}(c_V 
	+2d_V)\right)~,
	\label{lagmph}
\end{equation}
(iii) a photon-vector meson-pseudoscalar meson interactions,
\begin{eqnarray}
  &  & aL(V,\gamma,{\cal P}) = -\frac{a}{2}\frac{ie}{2}z^2_\pi + 
    A_\mu \left\{Tr(\{Q, [{\cal P},\partial^\mu {\cal P}]\}) + 
       c_V Tr(B\{Q, [{\cal P},\partial^\mu {\cal P}]\})\right.+
       \nonumber \\
  &  & \left. d_V \left( Tr(\{BQ, B[{\cal P},\partial^\mu {\cal P}]\})
       + 2Tr(BQ) Tr(B[{\cal P}, \partial^\mu {\cal P}])\right)\right\}
       +\ldots~,
       \label{lagphp}~,
\end{eqnarray}
and (iv) a vector-pseudoscalar meson interaction term,
\begin{eqnarray}
       &  &aL(V,{\cal P}) = -\frac{a}{2}\frac{gz^2_\pi}{2}
       \left\{Tr(\{V_\mu, [{\cal P},\partial^\mu {\cal P}]\}) + 
       c_V \left[Tr(B\{Q, [{\cal P},\partial^\mu {\cal P}]\})
       \right.\right. +
       \nonumber \\
       &  & \left.Tr(V_\mu) Tr (B[{\cal P},\partial^\mu {\cal P}])\right]
       \nonumber \\
 &  & \left. d_V \left[ Tr(\{BV_\mu, B[{\cal P},\partial^\mu {cal P}]\})
 + 2Tr(BV_\mu) Tr(B[{\cal P}, \partial^\mu {\cal P}]) \right]\right\} +
        \ldots~. 
	\label{lagvp}
\end{eqnarray}
Similarly, substituting the expansion of $\Delta_\mu$, Eqn.\ref{deltaf1}, into 
the pseudoscalar lagrangian part, $L_A + \bar{L}_A$, leads
to the photon-pseudoscalar meson interactions, 
\begin{equation}
	L(A,\gamma,{\cal P}) = -\frac{ie}{2}z^2_\pi A_\mu 
	\left\{Tr(\{[Q, {\cal P}],\partial^\mu {\cal P}\}) + 
	c_A Tr(B\{[Q, {\cal P}],\partial^\mu {\cal P}\})\right\}+\ldots~.
	\label{lagaphp}~.
\end{equation} 
Here the ellipsis  $\ldots$ denote terms involving 
$\partial{\cal P}{\cal P}^n,~~n\geq 2$. To summarize the photon-meson part
corresponding to the lagrangian, Eqn.\ref{elag} is given by,
\begin{equation}
	L_{VMDII} = -\frac{1}{4}Tr(V_{\mu\nu}V^{\mu\nu}) +
	aL(V,\gamma) + 	aL_{m\gamma} + aL(V,\gamma,{\cal P}) + 
	aL(V,{\cal P}) + L(A,\gamma,{\cal P})~.
	\label{lagvmd2}
\end{equation}
This last expression, with its photon mass $L_{m\gamma}$ term is rather 
similar to the popular VMDII model\cite{oconnell97}. By suitable 
transformation of the fields this term can be removed but prior to 
doing that some comments are in order. First, in complete analogy 
with the discussion of Ref.\cite{oconnell97}, we state here that the 
dressed photon propagator has the proper behavior of $-i/q^2$ at 
small photon momenta but significantly modified away from $q^2 = 0$.
 This certainly may be important in describing processes 
at sufficiently high momenta. Secondly, it is clear that the neutral 
$\rho^0, \omega$ and $\phi$ meson mix with the photon, spontaneously 
breaking symmetry down to the $U(1)_{em}$. We may exploit the 
diagonalization procedure described in section III to introduce 
$physical$ photon and mesons. In terms of these $physical$ fields, 
the lagrangian has no explicit coupling between the photon and neutral 
mesons, whilst there is a direct coupling between the photon to hadronic 
currents in the limit of exact $U(3)_F$. Thirdly, so far the value 
of the parameter $a$ is left free and in order to assign its value 
one needs considering the dynamics underlying QCD. Taking $a = 2$ 
at the instance of exact symmetry, $i.e. c_A = d_A = c_V = d_V = 0$, 
leads successfully to the phenomenological KSFR relations 
\cite{kawa66,riaz66}  
$m^2_\rho = 4F^2_8 g^2,\qquad g_\rho = 2 g_{\rho\pi\pi} F^2_\pi$ with 
$g_\rho~~\mbox{and}~~g_{\rho\pi\pi}$ are the 
$\rho\gamma~~\mbox{and}~~\rho\pi\pi$  effective coupling constants,
 respectively. Furthermore, with $a=2$ the symmetric 
lagrangian part $L(A,\gamma,{\cal P}) + aL(V,\gamma,{\cal P}) = 0$ so that 
the effective $\gamma\pi\pi$ coupling constant vanishes. Thus in the limit 
of exact symmetry the pion form factor is saturated by the $\rho$ meson 
contribution. All departures from the KSFR relations, nonvanishing 
effective $\gamma \pi \pi$ coupling $g_{\gamma\pi\pi} \neq 0$, and the 
$\rho$-meson dominance for the pion form factor are due to contributions 
from the symmetry breaking companions. The value $ a = 2.13 \pm 0.06$ 
deduced from the $\rho$ meson mass and its decay width into two pions, 
indicates that VMD provides, a good first approximation for the 
electromagnetic part of our lagrangian. With this value of $a$ our 
lagrangian includes in addition to the usual VMD corrections due to direct 
photon-pseudoscalar meson coupling.  

We now turn to derive a more elegant form of the lagrangian 
Eqn.\ref{lagvmd2} with manifestly a massless photon. We do that via 
transforming the vector meson fields into\cite{oconnell97}, 
\footnote{Note that the same way the momentum 
independent direct $V_\mu-Z^0_\mu~\mbox{and}~V_\mu-W_\mu$ couplings may 
be transformed into momentum dependent couplings}
\begin{equation}
	{\cal V}_\mu = V_\mu - \frac{e}{g}A_\mu Q~,
	\label{vectort}
\end{equation}
and define $\Gamma_\mu - g V_\mu = \hat{\Gamma}_\mu - g{\cal V}_\mu$ where 
$\hat{\Gamma}_\mu$ does not involve the $pure$ photon field contributions. 
In terms of these new fields, the lagrangian part $L_V +\bar{L}_V$ neither 
contains a photon mass term nor an explicit coupling of the photon to 
pseudoscalar currents, $j_\mu\sim [\partial_\mu {\cal P},{\cal P}]$. Thus  
the terms $L_{m\gamma}$ and $aL(V,\gamma,{\cal P})$ are removed from $L_V +
\bar{L}_V$. An explicit photon coupling to pseudoscalar currents, is now 
confined to the $L(A,\gamma,{\cal P})$ term resulting from the pseudoscalar 
$L_A + \bar{L}_A$ part which is left unaffected by the transformation 
Eqn.\ref{vectort}. Clearly, since the $\hat{\Gamma}_\mu$ as before involves 
terms like $-ieA_\mu({\cal P}Q{\cal P}-\{{\cal P}^2,Q\}/2)/2F^2_8$ etc. 
the lagrangian $L_V+\bar{L}_V$ still contains the $photon-vector~ meson 
-pseudoscalar~ meson$ couplings. Correspondingly, since
\begin{equation}
	V_{\mu\nu} = {\cal V}_{\mu\nu} -\frac{e}{g}A_{\mu\nu}Q~,
	\label{}
\end{equation}
the vector meson kinetic term transforms into,
\begin{equation}
	-\frac{1}{4}Tr(V_{\mu\nu}V^{\mu\nu}) = 
	- \frac{1}{4}\left(Tr({\cal V}_{\mu\nu}{\cal V}^{\mu\nu}) 
    -2\frac{e}{g}A_{\mu\nu}Tr(Q{\cal V}^{\mu\nu}) + 
    \frac{2}{3}\frac{e^2}{g^2} A_{\mu\nu}A^{\mu\nu}\right)~,
	\label{lagvkin}
\end{equation}
where as usual $A_{\mu\nu} = \partial_\mu A_\nu - \partial_\nu A_\mu$. 
The first term on rhs of Eqn. \ref{lagvkin} is the common kinetic term 
of the vector fields ${\cal V}$ while the second term  describes the direct 
photon-vector meson coupling. The last term is associated with the kinetic 
term of the photon $-A_{\mu\nu}A^{\mu\nu}/4$ and can be removed by 
rescalling the charge $e$ and the photon field according to\cite{oconnell97},
\begin{equation}
	 A'_{\mu\nu} = \sqrt{1+2e^2/3g^2} A_{\mu\nu};~~~~e' = 
	 e/\sqrt{1+2e^2/3g^2}~,
	\label{}
\end{equation}
Inserting all of these  modifications into the lagrangian 
Eqn.\ref{lagvmd2} 
leads to,
\begin{eqnarray}
	 &  & L_{VMDI}= - \frac{1}{4}\left(
	Tr({\cal V}_{\mu\nu}{\cal V}^{\mu\nu}) 
    -2\frac{e}{g}A_{\mu\nu}Tr(Q{\cal V}^{\mu\nu})\right) + 
	\nonumber \\
	 &  &  aF^2_8g^2\left[Tr({\cal VV}) 
	 +\tilde{w}_4Tr({\cal V})Tr({\cal V}) +
	 c_V\left(Tr(B{\cal VV}) +\tilde{w}_4Tr(B{\cal V})Tr({\cal V})\right) +
	d_V\left(Tr(B{\cal V}B{\cal V}) +
	  \tilde{w}_4 Tr(B{\cal V)}Tr(B{\cal V})\right)\right]
	\nonumber \\
	 &  &  -\frac{a}{2}\frac{gz^2_\pi}{2}
		\left\{Tr(\{{\cal V}_\mu, [{\cal P},\partial^\mu {\cal P}]\}) + 
	c_V \left[Tr(B\{Q, [{\cal P},\partial^\mu {\cal P}]\})\right.+
	Tr({\cal V}_\mu) Tr(B[{\cal P},\partial^\mu{\cal P}])\right]
	\nonumber \\
	 &  & \left. d_V\left[Tr(\{B{\cal V}_\mu,B[{\cal P},\partial^\mu{\cal 
P}]\})
	 + 2Tr(B{\cal V}_\mu) Tr(B[{\cal P},\partial^\mu{\cal P}])\right]\right\}
	 \nonumber \\
	 &   & -\frac{ie}{2}z^2_\pi A_\mu 
	\left\{Tr(\{[Q, {\cal P}],\partial^\mu {\cal P}\}) + 
	c_A Tr(B\{[Q, {\cal P}],\partial^\mu {\cal P}\})\right\}~,
	\label{lagvmd1}
\end{eqnarray}
where for convenience we have omitted the $'$ from $A'_\mu~\mbox{and}~e'$. 
The last expression is closely related to the VMDI model with the 
characteristics of (i) a massless photon, (ii) a momentum dependent
 direct photon-vector meson coupling ${\cal V}_{\mu\nu}A^{\mu\nu}$,
 and (iii) a direct photon-pseudoscalar meson coupling as well as the 
coupling through direct photon-vector meson transitions. 

To conclude this section we now use the lagrangian $L_{VMDI}$, 
Eq.\ref{lagvmd1}, to calculate the pion and kaon form factors. 
In the three diagram approximation (see Fig. 1) one obtains,
\begin{mathletters}
\begin{eqnarray}
  &  & F_\pi(q^2) = 1 +  q^2 \frac{a}{2}\frac{1}{m^2_\rho - q^2}~,
  \label{formfpi} \\
  &  & F_{K\pm}(q^2) = 1 +  q^2 
  \frac{3a}{2}z^2_K\left[\frac{1}{m^2_\rho - q^2}+
  \frac{1-c_V}{3}\frac{1}{m^2_\omega - q^2}+
  \frac{2(1+c_V+d_V)}{3}\frac{1}{m^2_\phi - q^2}\right]~, 
  \label{formfk} \\
   &  & F_{K^0}(q^2) = q^2
  \frac{3a}{2}z^2_K\left[\frac{1}{m^2_\rho - q^2}-
  \frac{1 - c_V}{3}\frac{1}{m^2_\omega - q^2} - 
   \frac{2(1+c_V+d_V)}{3}\frac{1}{m^2_\phi - q^2}\right]~, 
\label{formfk0}
\end{eqnarray}
\end{mathletters}
where $z_K = 1/\sqrt{1 + c_A /2}$ is a rescalling parameter defined from 
applying the diagonalization procedure to the pseudoscalar meson sector. 
It must be stressed that the three diagrams involving $V-A$ transition 
vertex contribute to those terms proportional to $Q^2$, only.
 The meson  charge radii are completely determined by diagrams 
involving the conversion of $\rho,~\omega,~\phi$ mesons into a
 photon (diagram 1b). We have
\begin{mathletters}
\begin{eqnarray}
	 &  & <r^2_\pi> = \frac{3a}{2}\frac{1}{m^2_\rho }~,
	\label{radpi} \\
	 &  & <r^2_{K\pm}> =
	\frac{3a}{2}z^2_K\left[\frac{1}{m^2_\rho }+
	\frac{1-c_V}{3}\frac{1}{m^2_\omega}+
	\frac{2(1+c_V+d_V)}{3}\frac{1}{m^2_\phi }\right]~, 
	\label{radk+} \\
	 &  & <r^2_{K0}> = 
	\frac{3a}{2}z^2_K\left[-\frac{1}{m^2_\rho}+
	\frac{1-c_V}{3}\frac{1}{m^2_\omega}+
	\frac{2(1+c_V+d_V)}{3}\frac{1}{m^2_\phi}\right]~.
	\label{radk0}
\end{eqnarray}
\end{mathletters}
The calculated radii of the pions and kaons along with  data are listed in 
Table \ref{tabradii}. In the first three columns we list the results 
from our model for the $Alternative ~I$ ($c_A=0.64\pm 0.06$),  
$Alternative ~II$ ($c_A=0.20\pm 0.05$) and QFB scheme ($c_A=1.4\pm 0.1$)
\cite{gedalin011}. The results from the $Alternative ~I$ and $Alternative 
~II$ explain equally well the pion and kaon radii although the
 $\chi^2$ criteria is slightly better for the former. The results from 
 the QFB scheme are by far inferior. For comparison we show the radii 
calculated with $a = 2.4$ as determined by Benayoun and 
O'Connell\cite{benayoun981}. Clearly, this value 
can not explain the charged pion radius.

\section{Summary and conclusions}

We have considered the vector meson sector within the framework of an 
effective field theory based on local $U(3)_L\bigotimes U(3)_R$ 
symmetry, where the vector mesons play the role of the dynamical 
gauge bosons of the HLS nonlinear chiral lagrangian. This supplement a 
previous study of the pseudoscalar sector\cite{gedalin011} within  
this same framework. The lagrangian is written in the most general 
way, including additional terms quadratic in the traces of the 
field matrices, which no longer vanish once singlet contributions 
are included. With these terms included, particularly the contribution 
from the small 
$\tilde{W}_4(X)Tr(\Gamma_\mu - g V_\mu)Tr(\Gamma_\mu - g V_\mu)$ term 
and its symmetry breaking companion, it is possible to reproduce the 
$\omega-\rho$ mass splitting. The explicit form of the lagrangian, through 
its kinetic and mass terms corresponding to vector mesons, determine the 
vector meson mixing scheme unambiguously. This observation which has been 
overlooked in previous studies allows us to calculate the model 
symmetry breaking scales and the value of $ag^2$ through the 
diagonalization of the mass matrix. Once the value of $g$ and $a$
 are disentangled, say from using another data point like the decay 
width of $\rho \to \pi \pi$, the lagrangian is completely defined, 
and can be used to calculate all observables without any additional 
assumption. As an example we have calculated the form factors and radii 
of the pions and kaons. Our calculations reproduce straightforwardly 
the proper normalization of the form factors and explain nicely 
the charged radii. We finally indicate that VMD model emerges naturally 
as a first approximation of our lagrangian. This stands in marked 
difference with Ref.\cite{benayoun981,harada01} 
where  VMD seems to be  badly violated.

\bigskip

{\bf Acknowledgment}  This work was supported in part by the Israel
Ministry of Absorption.

 \section{Appendix}
For convenience  we quote here the axial and vector parts of the 
lagrangian \ref{elag}, explicitly. First the symmetric parts are, 	
\begin{eqnarray}
     && L_{A}= W_{1}(X)Tr(\Delta_\mu\Delta^\mu) 
	+W_{4}(X)Tr(\Delta_\mu)Tr(\Delta^\mu) +
	 \nonumber\\
	 && W_{5}(X)Tr(\Delta_\mu)D^\mu \vartheta +
	 W_{6}(X)D_\mu \vartheta D^\mu \vartheta ~,
	\label{lagina}\\
	 &  & L_V =
	 \tilde{W}_{1}(X)Tr([\Gamma_\mu - gV_\mu][\Gamma^\mu - gV^\mu])+
	 \nonumber\\ 
	 &  & \tilde{W}_{4}(X)Tr(\Gamma_\mu - gV_\mu)Tr(\Gamma^\mu -
     gV^\mu)~,
	 \label{laginv}
\end{eqnarray}
where
\begin{eqnarray}
	 &  & 	D_\mu \vartheta = \partial_\mu \vartheta + Tr(r_\mu - l_\mu)~,
	\\
	 &  & V_{\mu\nu} = \partial_\mu V_\nu - \partial_\nu V_\mu -ig 
	 [V_\mu,V_\nu] ~.
\label{vmunu}
\end{eqnarray}
The symmetry breaking companions $\bar{L}_A$ are constructed in 
two alternative ways. The first (referred to as $Alternative ~I$) breaks 
$SU(3)_F$ symmetry, and corresponds to a quadratic form of the 
Goldstone meson kinetic energy term,
\begin{eqnarray}
         & &\bar{L}_A =
         W_1(X)\left(c_A Tr (B\bar{\Delta}_\mu
         \bar{\Delta}^\mu) + d_A Tr(B \bar{\Delta}_\mu
         B\bar{\Delta}^\mu)\right)+ \nonumber \\
         && W_4(X)d_A Tr(B\bar{\Delta}_\mu) Tr(B\bar{\Delta}^\mu)+
            W_5(X)c_A Tr(B \bar{\Delta}_\mu)D^\mu \vartheta ~.
        \label{labreak}
\end{eqnarray}
where the axial vector covariant,
\begin{equation}
	 \bar{\Delta}_\mu = 
	 \frac{i}{2}\left\{\xi^\dagger_8 ,\partial_{\mu}\xi_8\right\}
 +\frac{1}{2}\left(\xi^\dagger_8 r_\mu\xi_8 - 
	  \xi_8 l_\mu\xi^\dagger_8\right) ~,
	\label{bdelta} 
\end{equation}
involves the pseudoscalar octet field matrix only. The second alternative 
($Alternative~II$) breaks $U(3)_F$ symmetry and corresponds to bilinear 
kinetic energy term. Namely, 
 \begin{eqnarray}
     &&\bar{L}_{A}=
     W_{1}(X)\left(c_ATr(B\Delta_\mu\Delta^\mu)
         + d_ATr(B\Delta_\mu B\Delta^\mu)\right)
      \nonumber \\
         &&+W_{4}(X)\left(c_ATr(B\Delta_\mu)Tr(\Delta^\mu) +
         d_ATr(B\Delta_\mu)Tr(B\Delta^\mu)\right) +
         \nonumber\\
         && W_{5}(X)c_ATr(B\Delta_\mu)D^\mu \vartheta ~,
        \label{labreak1}
        \end{eqnarray}
with $\Delta_\mu$ involves now the nonet pseudoscalar field matrix. 
Similarly the asymmetric companion of ${L}_V$ is,
 \begin{eqnarray}
  &  & \bar{L}_V =
	 \tilde{W}_1(X)\left(c_V Tr (B[\Gamma_\mu -gV_\mu] 
	[\Gamma^\mu -gV^\mu])\right. +
	\nonumber\\
	&& \left. d_V Tr (B[\Gamma_\mu-gV_\mu] B [\Gamma^\mu -gV^\mu])\right) +
	 \nonumber\\
	 &&\tilde{W}_4(X)\left(c_VTr(\Gamma_\mu - gV_\mu)Tr(B[\Gamma^\mu -
           gV^\mu])\right.
	 \nonumber\\
	&& +\left. d_VTr(B[\Gamma_\mu - gV_\mu])Tr(B[\Gamma^\mu -
         gV^\mu])\right) ~.    
	\label{lvbreak}
 \end{eqnarray}	
In the expressions above, $c_A, d_A, c_V~\mbox{and}~d_V$ are the
symmetry breaking  parameters determined from data. In 
Ref.\cite{gedalin011} $c_A, d_A ~\mbox{and}~r =F_8/F_0$ were deduced from 
global fit of calculated radiative decay widths. Their values as 
determined  
for the {\it Alternative I\/}, {\it Alternative II\/} and QFB mixing 
schemes are listed in Table \ref{fit}.

\begin{table}[h]
	\caption{Pseudoscalar meson squared charge radii in $fm^2$}
	\begin{tabular}{cccccc}
		\hline
		 & {\it Alternative\/} I & {\it Alternative\/} II & QFB & 
		$g=2.4$ &Data  \\
		\hline
		$<r^2_{\pi^\pm}>$ & $0.43\pm 0.01$ & $0.43\pm 0.01$ & $0.43\pm 0.01$ 
		&$0.49\pm0.01$& $0.439\pm 0.03$  \\
		$<r^2_{K^\pm}>$ & $0.29\pm 0.02$ & $0.35\pm 0.02$ & $0.23\pm 0.01$ 
		&$0.33\pm0.02$& $0.31\pm 0.05$  \\
	    $<r^2_{k^0}>$ & $-0.040\pm 0.003$ & $-0.048\pm 0.003$ & $-0.031\pm 
0.003$ 
	    &$-(0.045\pm0.003)$ & $-0.054\pm 0.026$ \\
	   $\chi^2$ & $0.5$ &$0.7$ & $3.3$ & $2.1$ &  \\
	\end{tabular}
\protect\label{tabradii}
\end{table}

\begin{table}[h]
\caption{Symmetry breaking scales and $\chi^2/dof$ from global fit
to data. Values marked with an asterisk were kept fixed.  } 
 \begin{tabular}{ccccc}
 \hline
 & $c_W$ & $c_A$ & $d_A$ & $r$  \\
 \hline
 $Alternative~ I$ &$-(0.20\pm0.05)$ &$(0.64\pm0.06)$
 &$-0.25\pm0.04$ & $0.91\pm0.04$   \\
  $Alternative ~II$&$-(0.27\pm0.05)$&
$0.2\pm0.05$&$0.1\pm0.02$&$0.94\pm0.05$\\
  QFB &$-(0.19\pm0.05)$ &$(1.4\pm0.1)$ &$-1.1\pm0.1$ &
  *$1$  \\
 \end{tabular}
\protect\label{fit}
\end{table}

\begin{figure}
\includegraphics[scale=0.5]{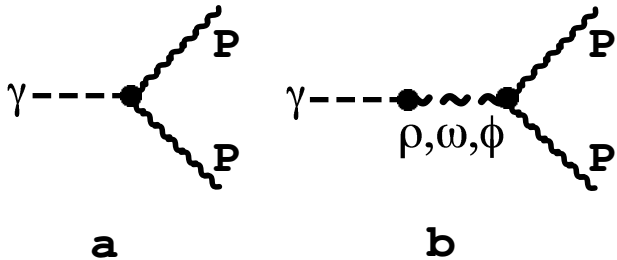}
\caption{ Diagram contributing into pseudoscalar meson form factors. 
Diagram $a$ corresponds to direct photon-pseudoscalar meson coupling, 
diagram $b$ involves the photon-vector meson conversion.
}
\label{fig1}
\end{figure}

\end{document}